\documentclass[a4paper]{jpconf}
\usepackage{graphicx}

\def\bea{\begin{eqnarray}}
\def\eea{\end{eqnarray}}
\def\beq{\begin{equation}}
\def\eeq{\end{equation}}

\def\be{\begin{equation}}
\def\ee{\end{equation}}

\newcommand{\ba}{\begin{eqnarray}}
\newcommand{\ea}{\end{eqnarray}}

\newcommand{\bml}{\begin{mathletters}}
\newcommand{\eml}{\end{mathletters}}

\def\sci#1#2{#1\times10^{#2}}

\def\fhatzero{\hat{f}_0}

\begin{document}
\title{Report on an all-sky LIGO search for periodic gravitational waves in the S4 data}

\author{Alicia M Sintes for the LIGO Scientific Collaboration}

\address{Departament de F\'{\i}sica, Universitat de les Illes
  Balears, Cra. Valldemossa Km. 7.5, E-07122 Palma de Mallorca,
  Spain}
\address{Max-Planck-Institut f\"ur
    Gravitationsphysik, Albert-Einstein-Institut, Am M\"uhlenberg 1,
    D-14476 Golm, Germany}
\ead{sintes@aei.mpg.de}

\begin{abstract}
We report on an all-sky search with the LIGO detectors for periodic 
gravitational waves in the frequency range 50-1000 Hz and having a negative 
frequency time derivative with magnitude between zero and $10^{-8}$ Hz/s. 
Data from the fourth LIGO science run  have been used in this search. 
Three different semi-coherent methods of summing strain power were applied.
Observing no evidence for periodic gravitational radiation, we report upper 
limits on strain amplitude and interpret these limits to constrain radiation 
from rotating neutron stars.
\end{abstract}

\section{Introduction}

The LIGO detector network consists of a 4-km interferometer in
Livingston Louisiana (called L1) and two interferometers in Hanford
Washington, one 4-km and another 2-km (H1 and H2, respectively) \cite{ligo1,ligo2}. 
The data analyzed in this paper were produced during LIGO's 29.5-day
fourth science run (S4)~\cite{S4InstrPaper,S4CalibrationNote}. This run started at noon 
Central Standard Time (CST) on February 22 and ended
at midnight CST on March 23, 2005. During the run, all three LIGO
detectors had displacement spectral amplitudes near 
$\sci{2.5}{-19}~{\rm m}~{\rm Hz}^{-1/2}$ in their most sensitive frequency band near 150 Hz.
In units of gravitational-wave strain amplitude, the sensitivity of H2
is roughly a factor of two worse than that of H1 and L1 over
much of the search band.

In this paper we summarize the results  from our search 
for periodic gravitational waves in the S4 data. 
The search was carried out
in the frequency range 50-1000 Hz, having a negative 
frequency time derivative with magnitude between zero and $1\times10^{-8}$ Hz/s
and over the entire sky. Isolated neutron stars in our galaxy 
were the prime target. For further details on the analysis and discussions 
we refer the reader to \cite{S4IncoherentPaper}.

\section{Overview of the search methods}

\begin{table}[h]
\caption{\label{ex} Summary of similarities and differences among the three
analysis methods used.}
\begin{center}
\begin{tabular}{llll}
\br
           & StackSlide	&Hough	&PowerFlux\\
\mr
Windowing	& Tukey	& Tukey	& Hann\\
Noise estimation & Median-based 	& Median-based 	&Time/frequency\\
   & floor tracking	& floor tracking	&decomposition\\
Line handling	& Cleaning	& Cleaning	& Skyband exclusion\\
Antenna pattern weighting	& No & 	Yes& 	Yes\\
Noise weighting &	No	& Yes	& Yes\\
Spindown step size&	$2 \times 10^{-10}$ Hz/s	& $2\times 10^{-10}$ Hz/s	&Frequency dependent\\
Limit at every skypoint	 & No	& No & 	Yes\\
Upper limit type	& Population-based	& Population-based	& Strict frequentist\\
\br
\end{tabular}
\end{center}
\end{table}

Three different analysis methods were considered for the search of periodic gravitational
wave signals. 
These have many features in common, but also have important differences that are 
summarized in Table~\ref{ex}.
All three methods are based on searching for cumulative excess power from a 
hypothetical periodic signal by examining successive spectral estimates
based on Short Fourier Transforms (SFTs) of 30-minute intervals of 
the calibrated detector strain data.
The simplest method used, known
as ``StackSlide'' \cite{BCCS,BC00,StackSlideTechNote, cgk}, averages
normalized power (i.e., power divided by an estimate of the spectral density of the noise) 
from each SFT. 
In the ``Hough'' method  \cite{pss01,hough04, S2HoughPaper,hough05,hough06, Palomba2005,
badrisintes},
the sum is of weighted binary zeroes or ones,
with weighting based on antenna pattern and detector noise,
where an SFT contributes only if the power exceeds a normalized power threshold. 
 This scheme also allows for a multi-interferometer search.
The third method, known as ``PowerFlux'' 
\cite{PowerFluxTechNote, PowerFluxPolarizationNote}, 
is a variant of the StackSlide method in which the power is weighted before summing.
In both the Hough and PowerFlux methods, the weights are chosen according
to the noise and detector antenna pattern to maximize the signal-to-noise ratio.
  
  Each method corrects explicitly for sky-position dependent
Doppler modulations of the apparent source
frequency due to the Earth's rotation and its orbital motion 
with respect to the Solar System Barycenter (SSB),
 and the frequency's time derivative, intrinsic to
the source. 
This requires a search in a four-dimensional parameter space;
a template in the space refers to a set of values: 
${\mathbf \lambda} = \{ \fhatzero, \dot{f}, \alpha, \delta \}$.
The third method, PowerFlux, also searches explicitly
over polarization angle, so
that ${\mathbf \lambda} = \{ \fhatzero, \dot{f}, \alpha, \delta, \psi \}$.
 All three methods search for initial frequency $\fhatzero$ in the
range $50\,$--$\,1000$~Hz with a uniform grid spacing equal to 
the size of an SFT frequency bin.
The range of $\fhatzero$ is determined by the noise curves of the
interferometers, likely detectable source frequencies,
and limitations due to the increasing computational cost at
high frequencies. 
The range of $\dot{f}$ values searched is 
$[-\sci{1}{-8}$,~$\,0]~\mathrm{Hz}~\mathrm{s}^{-1}$
for the StackSlide and PowerFlux methods and 
$[-\sci{2.2}{-9}$,~$\,0]~\mathrm{Hz}~\mathrm{s}^{-1}$ for
the Hough method. The ranges of $\dot{f}$ are narrow enough to
complete the search in a reasonable amount of time, yet wide enough to
include likely signals.
The number of sky points
that must be searched grows quadratically with
the frequency $\fhatzero$, ranging in this search from about five thousand at
50 Hz to about two million at 1000 Hz. 
 All three methods use
nearly isotropic grids which cover the entire sky. 

Other differences among the methods 
concern the data windowing and filtering used in
computing Fourier transforms and data handling.
StackSlide and Hough apply high pass filters to the data above 
$40 {\rm Hz}$, in addition to the filter used to produce the
calibrated data stream, and 
use Tukey windowing. PowerFlux applies no additional filtering and uses Hann
windowing with 50\% overlap between adjacent SFT's. StackSlide
and Hough use median-based noise floor tracking. 
In contrast, Powerflux uses a time-frequency decomposition.
The StackSlide and Hough searches used 1004 SFTs from H1 and 899
from L1, the two interferometers with the best broadband sensitivity. 
For PowerFlux, the corresponding numbers of overlapped SFTs were 1925 and 1628. 
The Hough search also used 1063 H2 SFTs.
Sharp instrumental lines, which can mimic a continuous gravitational-wave
signal for parameter space points that correspond to small Doppler modulation,
are also handled differently.  StackSlide and Hough carry out
removal of known instrumental lines of varying widths in individual
SFTs. The measured powers in those bins are replaced with random
noise generated from the noise observed in neighboring bins.
  PowerFlux  flags single-bin lines  during
data preparation so that when searching for a particular source
an individual SFT bin power is ignored when it coincides with
the source's apparent frequency. 
The PowerFlux search
also divides the sky into regions according to susceptibility to stationary
instrumental line artifacts and excludes the less favorable ones when setting upper limits.

\section{Summary of results}

 \begin{figure}
\begin{center}
\includegraphics[width=15cm]{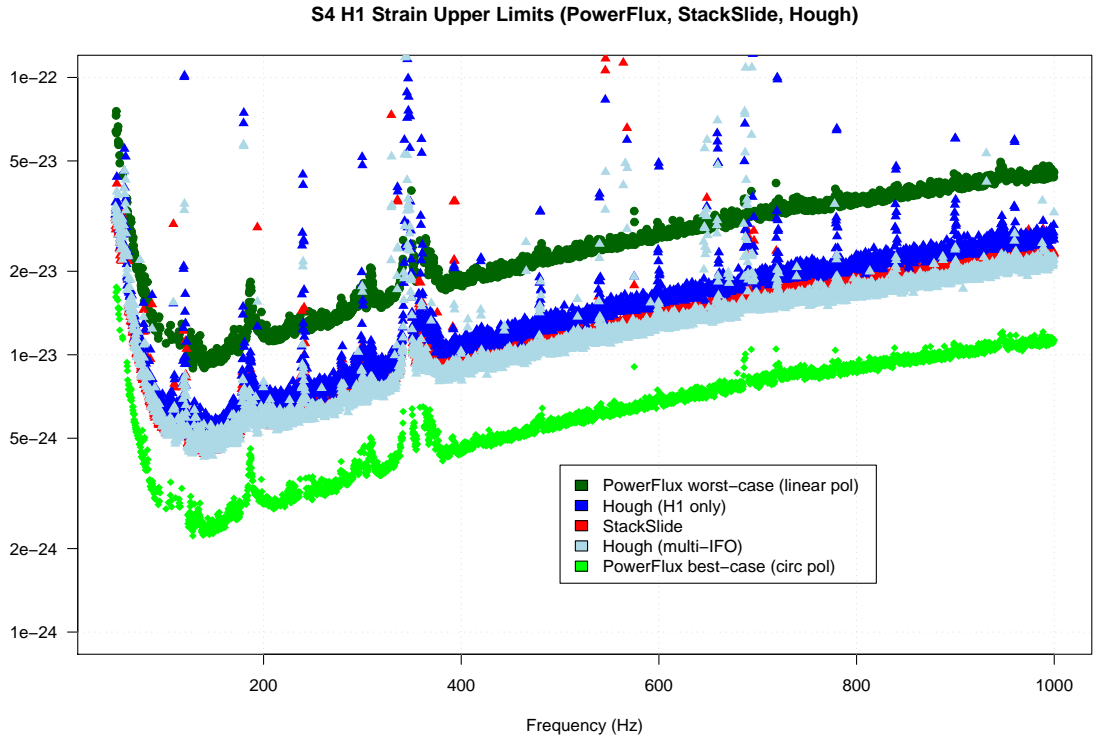}
\end{center}
\caption{\label{UL} H1 upper limits on $h_0$ from the three methods. The StackSlide and
   Hough limits are population-based, while those from PowerFlux are strict and apply,
   respectively, to the most favorable and least favorable pulsar inclinations. Also
   shown are the multi-interferometer limits from the Hough search.}
\end{figure}

All three methods described above have been applied
in an all-sky search over a frequency range 50-1000 Hz. 
All the loud candidates produced were checked for coincidences between H1 and L1
and the only
surviving candidates were associated with the hardware injected pulsars and instrumental lines.
Also different types of qualitative follow-up tests were performed on each of the coincident outliers.
As described in \cite{S4IncoherentPaper}, no evidence for a periodic
gravitational wave signal was observed in any of the searches and upper limits
on sources were determined. 

For the StackSlide and Hough methods, 
95\% confidence-level frequentist upper limits were placed
on putative rotating neutron stars, assuming a uniform-sky and isotropic-orientation
parent sample. Depending on the source location and inclination, these limits
may overcover or undercover the true 95\% confidence-level band. 
For the PowerFlux method, strict frequentist upper limits were placed on linearly and circularly
polarized periodic gravitational wave sources, assuming {\it worst-case}
sky location, avoiding undercoverage. The limits on linear polarization are also
re-interpreted as limits on rotating neutron stars, assuming worst-case sky location
and worst-case star inclination. 

 Figure \ref{UL} shows  superimposed the
final upper limits on $h_0$ from the StackSlide, Hough, and PowerFlux methods when applied
to the S4 single-interferometer H1  data, together with the multi-interferometer H1+H2+L1 Hough search.
The reader should notice that the Hough search
sensitivity improves with the summing of powers
from two or more interferometers. 
Over the LIGO frequency band of sensitivity,
these S4 all-sky upper limits are approximately  an order of 
magnitude better than those published
previously from the second science run (S2)~\cite{S2FstatPaper,S2HoughPaper}. 
The best population-based upper limit with $95\%$ confidence on the gravitational-wave
strain amplitude, found for simulated sources distributed isotropically across the sky
and with isotropically distributed spin-axes,
is $4.28 \times 10^{-24}$ (near 140 Hz) for the multi-interferometer  Hough search. 

In this search we have reached an important milestone on the road to astrophysically
interesting all-sky results:
Our best upper limits on $h_0$ are comparable to the value of a few times
$10^{-24}$ at which one might optimistically expect to see the strongest
signal from a previously unknown neutron star according to a generic
argument originally made by Blandford (unpublished) and extended in our
previous search for such objects in S2 data \cite{S2FstatPaper}.
Moreover, the multi-interferometer Hough transform search could have detected an
object at the distance of the nearest known neutron star RX~J1856.5$-$3754,
which is about 110--170~pc from Earth.

Comparing the three methods we found that Hough is computationally 
faster and more robust against large transient power artifacts,
but is slightly less sensitive than StackSlide for stationary data \cite{S2HoughPaper,StackSlideTechNote}. 
The PowerFlux method is found in most frequency ranges to have better detection efficiency than
the StackSlide and Hough methods, the exceptions occurring in bands with large non-stationary
artifacts, for which the Hough method proves more robust. However, the StackSlide and Hough methods
can be made more sensitive by starting with the maximum likelihood statistic
(known as the $\cal F$-statistic \cite{jks,hough04,S2FstatPaper}) rather than SFT power as
the input data, though this improvement comes at an increased computational cost.
The trade-offs among the methods means that each could play a role in our future searches.
 The lower computational cost of
the Hough search would be an advantage in this case. Multi-interferometer searches also increase
the sensitivity, while reducing outliers (false-alarms), without having to increase greatly
the size of the parameter space used, as illustrated by the Hough search in this paper.
A fifth science run (S5), which started in November 2005 and finished at the end of September
2007, has generated data at initial LIGO's design sensitivity. Our plans are to search this
data using the best methods possible, based on what is learned from this and previous
analyses, and given the computational bounds.

\ack

The authors gratefully acknowledge the support of the United States
National Science Foundation for the construction and operation of the
LIGO Laboratory and the Science and Technology Facilities Council of the
United Kingdom, the Max-Planck-Society, and the State of
Niedersachsen/Germany for support of the construction and operation of
the GEO600 detector. The authors also gratefully acknowledge the support
of the research by these agencies and by the Australian Research Council,
the Council of Scientific and Industrial Research of India, the Istituto
Nazionale di Fisica Nucleare of Italy, the Spanish Ministerio de
Educacion y Ciencia, the Conselleria d'Economia Hisenda i Innovacio of
the Govern de les Illes Balears, the Scottish Funding Council, the
Scottish Universities Physics Alliance, The National Aeronautics and
Space Administration, the Carnegie Trust, the Leverhulme Trust, the David
and Lucile Packard Foundation, the Research Corporation, and the Alfred
P. Sloan Foundation.

\end{document}